\documentclass[10pt,conference]{IEEEtran}
\IEEEoverridecommandlockouts
\usepackage{cite}
\usepackage{amsmath,amssymb,amsfonts}
\usepackage{algorithmic}
\usepackage{graphicx}
\usepackage{textcomp}
\usepackage{xcolor}
\usepackage{booktabs} % For formal tables
\usepackage{multirow}
\usepackage{xspace}
\usepackage{paralist}
\usepackage{enumitem}
\usepackage{booktabs,comment,flushend,listings,multirow,tikz,todonotes}
\usepackage{graphicx,caption,subcaption}
\usepackage{amssymb}
\usepackage{ifthen}
\usetikzlibrary{arrows,calc,shapes}
\usepackage{wrapfig}
\usepackage{xcolor}
\usepackage{listings}
\usepackage{xspace}

\def\BibTeX{{\rm B\kern-.05em{\sc i\kern-.025em b}\kern-.08em
    T\kern-.1667em\lower.7ex\hbox{E}\kern-.125emX}}

\colorlet{light-gray}{gray!20}
\definecolor{pblue}{rgb}{0.13,0.13,1}
\definecolor{pgreen}{rgb}{0,0.5,0}
\definecolor{pred}{rgb}{0.9,0,0}
\definecolor{pgrey}{rgb}{0.46,0.45,0.48}

\newcommand{\ie}{\textit{i.e.,}\xspace}
\newcommand{\eg}{\textit{e.g.,}\xspace}
\newcommand{\etc}{\textit{etc.}\xspace}
\newcommand{\etal}{\textit{et al.}\xspace}

\newcommand{\secref}[1]{Section~\ref{#1}\xspace}

\newcommand{\tabref}[1]{Table~\ref{#1}\xspace}

\begin{document}

\title{Learning How to Mutate Source Code\\from Bug-Fixes\vspace{0cm}}

\sloppy

\author{
	\IEEEauthorblockN{Michele~Tufano\IEEEauthorrefmark{1},~Cody~Watson\IEEEauthorrefmark{1},~Gabriele~Bavota\IEEEauthorrefmark{2},~Massimiliano~Di~Penta\IEEEauthorrefmark{3},~Martin~White\IEEEauthorrefmark{1},~Denys~Poshyvanyk\IEEEauthorrefmark{1}}
	\IEEEauthorblockA{\IEEEauthorrefmark{1}College~of~William~and~Mary,~Williamsburg,~Virginia,~USA\\Email:~\{mtufano,~cawatson,~mgwhite,~denys\}@cs.wm.edu}
	\IEEEauthorblockA{\IEEEauthorrefmark{2}Universit\`{a} della Svizzera italiana (USI),~Lugano,~Switzerland\\Email:~gabriele.bavota@usi.ch}
	\IEEEauthorblockA{\IEEEauthorrefmark{3}University of Sannio,~Benevento,~Italy\\Email:~dipenta@unisannio.it}
}

\maketitle

\begin{abstract}
Mutation testing has been widely accepted as an approach to guide test case generation or to assess the effectiveness of test suites. Empirical studies have shown that mutants are representative of real faults; yet they also indicated a clear need for better, possibly customized, mutation operators and strategies.
While methods to devise domain-specific or general-purpose mutation operators from real faults exist, they are effort- and error-prone, and do not help the tester to decide whether and how to mutate a given source code element.
We propose a novel approach to automatically learn mutants from faults in real programs. First, our approach processes bug fixing changes using fine-grained differencing, code abstraction, and change clustering. Then, it learns mutation models using a deep learning strategy. We have trained and evaluated our technique on a set of $\sim$787k bug fixes mined from GitHub.   Our empirical evaluation showed that our models are able to predict mutants that resemble the actual fixed bugs in between 9\% and 45\% of the cases, and
over 98\% of the automatically generated mutants are lexically and syntactically correct.
\end{abstract}

\begin{IEEEkeywords}
	mutation~testing,~%
deep~learning,~%
%language~models,~%
%code~similarities,~%
%machine~learning,~%
neural~networks
\vspace{-0.4cm}
\end{IEEEkeywords}

\section{Introduction}
\label{sec:intro}
% !TeX root = ../ms.tex

Mutation testing aims at injecting artificial faults into the
 program's source code or bytecode~\cite{Hamlet:TSE,DeMillo:Computer} to simulate defects.  Mutants (\ie versions of the program with an artificial defect) can guide the design or even the automatic generation of a test suite \cite{FraserA11,FraserA13}, and can be used to assess the effectiveness of an existing test suite~\cite{MouchawrabBLP11,BriandPL04}.
 
A number of studies have been dedicated to understand the interconnection between mutants and real faults  \cite{AndrewsBL05,AndrewsBLN06,DaranT96,Just:ISSTA14,Luo18,Shamshiri:ASE15,Just:FSE14,Chekam:ICSE'17}, indicating that carefully-selected mutants can be as effective as real faults
\cite{AndrewsBL05,AndrewsBLN06,DaranT96}, but also that mutants can underestimate a test suite's fault detection capability \cite{AndrewsBL05,AndrewsBLN06}. As pointed out by Just \etal \cite{Just:FSE14}, there is a need to improve mutant taxonomies in order to make them more representative of real faults.
 Taxonomies of mutants have been devised by taking typical bugs into account \cite{DBLP:journals/tosem/OffuttLRUZ96,DBLP:journals/stvr/KimCM01}, and even coping with specific domains
\cite{DBLP:conf/sigsoft/JabbarvandM17, wang2015mualloy, 7107450, Offutt:2001:MUO:571305.571314, VasquezBTMPVBP17}.  Brown \etal \cite{Brown:2017:CFW:3106237.3106280} leveraged bug-fixes to infer 7.5k types of mutation operators from diff patches. However, devising specific mutant taxonomies  requires a substantial manual effort, and may not generalize across projects.

Stemming from the previous considerations by 
 Brown \etal \cite{Brown:2017:CFW:3106237.3106280}, as well as from recent work aimed at learning bug repairs from an existing set of previous fixes \cite{TufanoWBPWP18,abs-1812-08693} and, more generally, from the successful applications of machine learning on code to support several SE tasks tasks \cite{WangLT16,WhiteTVP16,DBLP:conf/iwpc/LamNNN17,DBLP:conf/sigsoft/GuZZK16, DeepCodeSearch, Raychev:2014:CCS:2594291.2594321, Hellendoorn:2017:DNN:3106237.3106290, DBLP:journals/corr/abs-1711-00740, allamanis2016learning, Allamanis:2015:SAM:2786805.2786849}, we conjecture that \emph{mutants can be automatically learned from previous fixes}. We propose an approach for automatically learning mutants from actual bug fixes. After having mined bug-fixing commits from software repositories, we extract change operations using an AST-based differencing tool and abstract them. Then, to enable the learning of specific mutants, we cluster similar changes together. Finally, we learn from the changes using a Recurrent Neural Network (RNN) Encoder-Decoder architecture \cite{kalchbrenner-blunsom:2013:EMNLP, DBLP:journals/corr/SutskeverVL14, DBLP:journals/corr/ChoMGBSB14}. When applied to unseen code, the learned model decides in which location and what changes should be performed. Besides being able to learn mutants from an existing source code corpus, and differently from Brown \etal \cite{Brown:2017:CFW:3106237.3106280}, our approach is also able to determine where and how to mutate source code, as well as to introduce new literals and identifiers in the mutated code. 

A buggy code fragment arguably represents the perfect mutant for the fixed code because: (i) it is a mutation of the fixed code; (ii) such a mutation exposed a buggy behavior; (iii) the buggy code does not represent a trivial mutant; (iv) the test suite did not detect the bug in the buggy version. 

We evaluate our approach on $787k$ bug-fixing commits with the aim of investigating (i) how similar the learned mutants are as compared to real bugs; (ii) how specialized models (obtained by clustering changes) can be used to generate specific sets of mutants; and (iii) from a qualitative point of view, what operators were the models able to learn. The results indicate that our approach is able to generate mutants that perfectly correspond to the original buggy code in 9\% to 45\% of cases (depending on the model). Most of the generated mutants are syntactically correct (more than 98\%), and the specialized models are able to inject different types of mutants. 

This paper provides the following contributions:
\begin{itemize}[noitemsep,wide=0pt, leftmargin=\dimexpr\labelwidth + 0\labelsep\relax]
	\item A novel approach for learning how to mutate code from bug-fixes. 
	\item Empirical evidence that our models are able to learn diverse mutation operators that are closely related to real bugs.
	\item Data and source code to enable replication \cite{online-appendix}.
\end{itemize}
\vspace{-0.2cm}

\section{Approach}
\label{sec:design}
% !TeX root = ../ms.tex

We start by mining bug-fixing commits from thousands of GitHub repositories (Sec. \ref{sec:mining}). From the bug-fixes, we extract method-level pairs of \textit{buggy} and corresponding \textit{fixed} code that we call \textit{transformation pairs} (TPs) (Sec. \ref{sec:tp_extraction}). 

TPs represent the examples we use to learn how to mutate code from bug-fixes (\textit{fixed} $\rightarrow$ $buggy$). We rely on GumTree \cite{DBLP:conf/kbse/FalleriMBMM14} to extract a list of edit actions ($A$) performed between the buggy and fixed code. Then, we use a Java Lexer and Parser to abstract the source code of the TPs (Sec. \ref{sec:tp_abstraction}) into a representation that is more suitable for learning.
The output of this phase is the set of abstracted TPs and their corresponding mapping $M$ which allows reconstructing the original source code. Next, we generate different datasets of TPs (Sec. \ref{sec:ident_lit} and \ref{sec:clustering}). Finally, for each set of TPs we use an encoder-decoder model to learn how to transform a \textit{fixed} piece of code into the corresponding \textit{buggy} version (Sec. \ref{sec:learning}).

\subsection{Bug-Fixes Mining}
\label{sec:mining}
We downloaded the GitHub Archive \cite{githubarchive} containing every public GitHub event between March 2011 and October 2017. Then, we used the Google BigQuery APIs to identify commits related to bug-fixes. We selected all the commits having a message containing the patterns:  (``fix'' or ``solve'') and (``bug'' or ``issue'' or ``problem'' or ``error'').
We identified 10,056,052 bug-fixing commits for which we extracted the commit ID (SHA), the project repository, and the commit message. 

Since not all commits matching our pattern are necessarily related to corrective maintenance \cite{AntoniolAPKG08, HerzigJZ13}, we assessed the precision of the regular expression used to identify bug-fixing commits. Two authors independently analyzed a statistically significant sample (95\% confidence level $\pm5\%$ confidence interval, for a total size of 384) of identified commits to judge whether the commits were actually referring to bug-fixing activities. Next, the authors met to resolve a few disagreements in the evaluation (13 cases). The evaluation results reported a true positive rate of 97\% \cite{online-appendix}. The commits classified as false positives mainly referred to partial/incomplete fixes.

After collecting the bug-fixing commits, for each commit we extracted the source code pre- and post- bug-fixing (\ie buggy and fixed code) using the GitHub Compare API \cite{github-compare}. 
We discarded (i) files created in the bug-fixing commit since there is no buggy version to learn from; (ii)  commits that had touched more than five Java files, since we aim to learn from bug-fixes focusing on only a few files and not spread across the system; and (iii) large commits that are more likely to represent tangled changes \cite{HerzigZ13}, \ie dealing with different tasks. Also, we excluded commits related to repositories written in languages different than Java, since we aim at learning mutation operators for Java code. After these filtering steps, we extracted the pre- and post-code from 787,178 bug-fixing commits.

\subsection{Analysis of Transformation Pairs}
A TP is a pair $(m_b, m_f)$ where $m_b$ represents a buggy code component and $m_f$ represents the corresponding fixed code. We will use these TPs as examples when training our RNN. The idea is to train the model to learn the transformation from the fixed code component ($m_f$) to the buggy code ($m_b$), in order to generate mutants that are similar to real bugs.

\subsubsection{Extraction}
\label{sec:tp_extraction}
Given a bug-fix \textit{bf}, we extracted the buggy files ($f_b$) and the corresponding fixed ($f_f$) files. For each pair $(f_b, f_f)$, we ran AST differencing between the ASTs of $f_b$ and $f_f$ using GumTree Spoon AST Diff \cite{DBLP:conf/kbse/FalleriMBMM14}, to compute the sequence of AST edit actions that transforms $f_b$ into $f_f$.

Instead of computing the AST differencing between the entire buggy and fixed files, we separate the code into method-level pieces that will constitute our TPs. We first rely on GumTree to establish the mapping between the nodes of $f_b$ and $f_f$. Then, we extract the list of mapped pairs of methods $L = \{(m_{1b}, m_{1f}), \dots , (m_{nb}, m_{nf}) \}$. Each pair $(m_{ib}, m_{if})$ contains the method $m_{ib}$ (from the buggy file $f_b$) and the corresponding mapped method $m_if$ (from the fixed file $f_f$). Next, for each pair of mapped methods, we extract a sequence of edit actions using the GumTree algorithm. We then consider only those method pairs for which there is at least one edit action (\ie we disregard methods unmodified during the fix). Therefore, the output of this phase is a list of $TPs = \{tp_1, \dots, tp_k\}$, where each TP is a triplet $tp = \{m_b, m_f, A\}$, where $m_b$ is the buggy method, $m_f$ is the corresponding fixed method, and $A$ is a sequence of edit actions that transforms $m_b$ in $m_f$. We do not consider any methods that have been newly created or completely deleted within the fixed file since we cannot learn mutation operations from them. Also, TPs do not capture changes performed outside methods (\eg class name).

The rationale behind the choice of method-level TPs is manyfold. First, methods represent a reasonable target for mutation, since they are more likely to implement a single task. Second, methods provide enough meaningful context for learning mutations, such as variables, parameters, and method calls used in the method. Smaller snippets of code lack the necessary context. Third, file- or class-level granularity could be too large to learn patterns of transformation. Finally, considering arbitrarily long snippets of code, such as hunks in diffs, could make the learning more difficult given the variability in size and context \cite{10.1007/978-3-642-35843-2_6,DBLP:conf/iwpc/AlaliKM08}. Note that we consider each TP as an independent fix, meaning that multiple methods modified in the same bug fixing activity are considered independently from one other. 
In total, we extracted $\sim$2.3M TPs.

\subsubsection{Abstraction}
\label{sec:tp_abstraction}
The major problem in dealing with raw source code in TPs is the extremely large vocabulary created by the multitude of identifiers and literals used in the code of the $\mathtt{\sim}$2M mined projects. This large vocabulary would hinder our goal of learning transformations of code as a neural machine translation task. Therefore, we abstract the code and generate an expressive yet vocabulary-limited representation. We use a combination of a Java lexer and parser to represent each buggy and fixed method within a TP, as a stream of tokens. First, the lexer (based on ANTLR \cite{Parr:2011:LFA:1993498.1993548, Parr:2013:DAR:2501720}) reads the raw code tokenizing it into a stream of tokens. The tokenized stream is then fed into a Java parser \cite{javaparser}, which discerns the role of each identifier (\ie whether it represents a variable, method, or type name) and the type of literals. Each TP is abstracted in isolation. Given a TP $tp=\{m_b, m_f, A\}$, we first consider the source code of $m_f$. The source code is fed to a Java lexer, producing the stream of tokens. The stream of tokens is then fed to a Java parser, which recognizes the identifiers and literals in the stream. The parser then generates and substitutes a unique ID for each identifier/literal within the tokenized stream. If an identifier or literal appears multiple times in the stream, it will be replaced with the same ID. The mapping of identifiers/literals with their corresponding IDs is saved in a map ($M$). The final output of the Java parser is the abstracted method ($abstract_f$). Then, we consider the source code of $m_b$. The Java lexer produces a stream of tokens, which is then fed to the parser. The parser continues to use map $M$ for $m_b$. The parser generates new IDs only for novel identifiers/literals, not already contained in $M$, meaning, they exist in $m_b$ but not in $m_f$. Then, it replaces all the identifiers/literals with the corresponding IDs, generating the abstracted method ($abstract_b$). The abstracted TP is now the following 4-tuple $tpa = \{abstract_b, abstract_f, A, M\}$, where $M$ is the ID mapping for that particular TP. The process continues considering the next TP, generating  a  new mapping $M$. Note that we first analyze the fixed code $m_f$ and then the corresponding buggy code $m_b$ of a TP since this is the direction of the learning process (from $m_b$ to $m_f$).

\begin{figure}[b]
\vspace{-0.75cm}
	\centering
	\includegraphics[width=0.9\linewidth]{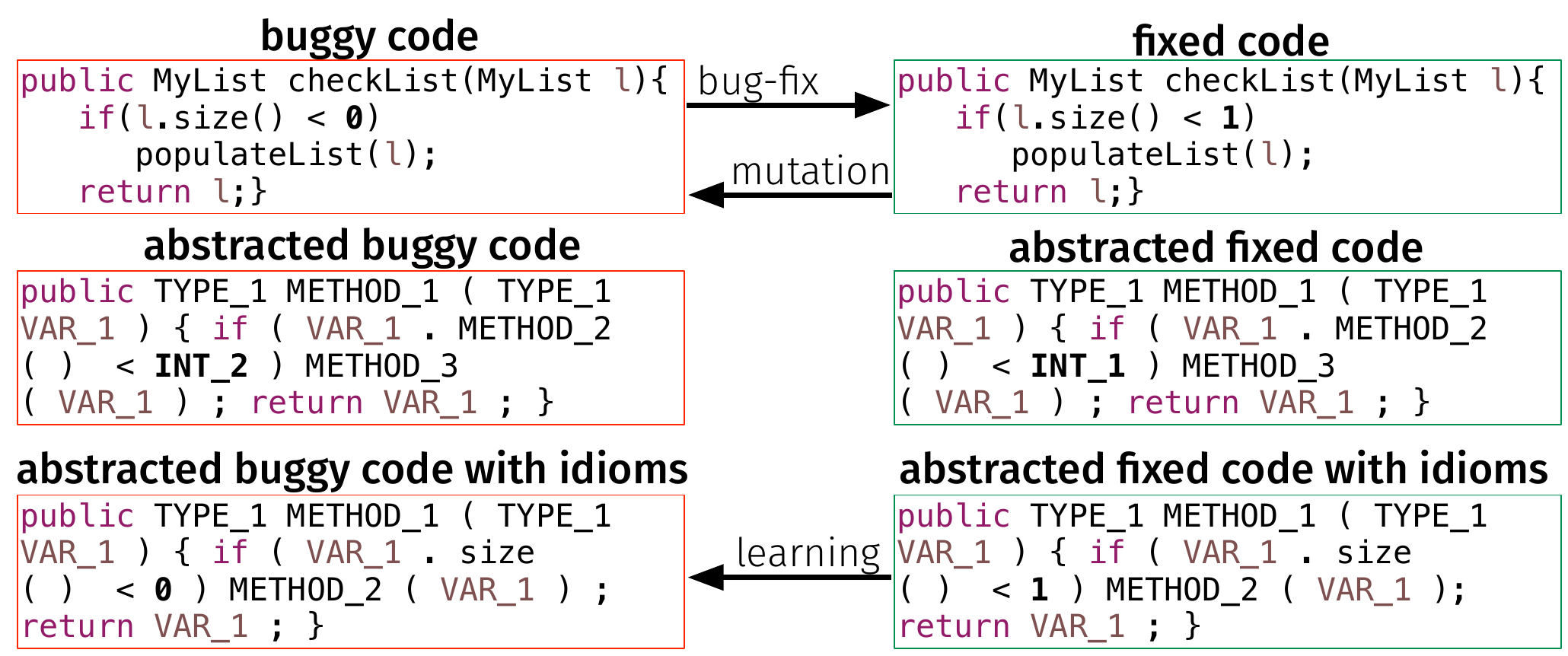}
	\vspace{-0.1cm}
	\caption{Transformation Pair Example.}
	\label{fig:tp}
\end{figure}

The assignment of IDs to identifiers and literals occurs in a sequential and positional fashion. Thus, the first method name found will receive the ID \texttt{METHOD\_1}, likewise the second method name will receive ID \texttt{METHOD\_2}. This process continues for all method and variable names (\texttt{VAR\_X}) and literals (\texttt{STRING\_X}, \texttt{INT\_X}, \texttt{FLOAT\_X}). Fig. \ref{fig:tp} shows an example of the TP's abstracted code. It is worth noting that IDs are shared between the two versions of the methods and new IDs are generated only for newly found identifiers/literals. The abstracted code allows to substantially reduce the number of unique words in the vocabulary because we allow the reuse of IDs across different TPs. For example, the first method name identifier in any transformation pair will be replaced with the ID \texttt{METHOD\_1}, regardless of the original method name. 

At this point, $abstract_b$ and $abstract_f$ of a TP are a stream of tokens consisting of language keywords (\eg \texttt{if}), separators (\eg ``('', ``;'') and IDs representing identifiers and literals. Comments and annotations have been removed.

Fig. \ref{fig:tp} shows an example of a TP. The left side is the buggy code and the right side is the same method after the bug-fix (changed the $\mathtt{if}$ condition). The abstracted stream of tokens is shown below each corresponding version of the code. Note that the fixed version is abstracted before the buggy version. The two abstracted streams share most of the IDs, except for the \texttt{INT\_2} ID (corresponding to the int value 0), which appears only in the buggy version. 

There are some identifiers and literals that appear so often in the source code that, for the purpose of our abstraction, they can almost be treated as keywords of the language. For example, the variables \texttt{i}, \texttt{j}, or \texttt{index} are often used in loops. Similarly, literals such as \texttt{0}, \texttt{1}, \texttt{-1} are often used in conditional statements and return values. Method names, such as \texttt{getValue}, appear multiple times in the code as they represent common concepts. These identifiers and literals are often referred to as ``idioms'' \cite{Brown:2017:CFW:3106237.3106280}.  We keep these idioms in our representation, that is, we do not replace idioms with a generated ID, but rather keep the original text in the code representation. To define the list of idioms, we first randomly sampled 300k TPs and considered all their original source code. Then, we extracted the frequency of each identifier/literal used in the code, discarding keywords, separators, and comments. Next, we analyzed the distribution of the frequencies and focused on the top frequent words (outliers of the distribution). In particular, we focused on the top $0.005\%$ of the distribution. Two authors manually analyzed this list and curated a set of 272 idioms. Idioms also include standard Java types such as \texttt{String}, \texttt{Integer}, common \texttt{Exceptions}, \etc The complete list of idioms is available on our online appendix \cite{online-appendix}.

Fig. \ref{fig:tp} shows the \textit{idiomized} abstracted code at the bottom. The method name \texttt{size} is now kept in the representation and not substituted with an ID. This is also the case for the literal values \texttt{0} and \texttt{1}, which are very frequent idioms. Note that the method name \texttt{populateList} is now assigned ID \texttt{METHOD\_2} rather than \texttt{METHOD\_3}.
This representation provides enough context and information to effectively learn code transformations, while keeping a limited vocabulary ($|V| = \mathtt{\sim}$430). Note that the abstracted code can be mapped back to the real source code using the mapping ($M$).

\subsubsection{Filtering Invalid TPs}

Given the extracted list of 2.3M TPs, we manipulated their code via the aforementioned abstraction method. During the abstraction, we filter out such TPs that: (i) contain lexical or syntactic errors (\ie either the lexer or parser failed to process them) in either the buggy or fixed version of the code; (ii) their buggy and fixed abstracted code ($abstract_b$, $abstract_f$) resulted in equal strings. The equality of $abstract_b$ and $abstract_f$ is evaluated while ignoring whitespace, comments or annotations edits, which are not useful in learning mutants. Next, we filter out TPs that performed more than 100 atomic AST actions ($|A|>100$) between the buggy and fixed version. The rationale behind this decision was to eliminate outliers of the distribution (the 3rd quartile of the distribution is 14 actions) which could hinder the learning process. Moreover, we do not aim to learn such large mutations. Finally, we discard long methods and focus on small/medium size TPs. We filter out TPs whose fixed or buggy abstracted code is longer than 50 tokens. We discuss this choice in the \secref{sec:threats} and report preliminary results also for longer methods. After the filtering, we obtained  $\mathtt{\sim}$380k TPs.

\subsubsection{Synthesis of Identifiers and Literals}
\label{sec:ident_lit}

TPs are the examples we use to make our model learn how to mutate source code. Given a $tp = \{m_b, m_f, A\}$, we first abstract its code, obtaining $tpa = \{abstract_b, abstract_f, A, M\}$. The fixed code $abstract_f$ is used as input to the model which is trained to output the corresponding buggy code (mutant) $abstract_b$. This output can be mapped back to real source code using $M$.

In the current usage scenario (\ie generating mutants), when the model is deployed, we do not have access to the oracle (\ie buggy code, $abstract_b$), but only to the input code. This source code can then be abstracted and fed to the model, which generates as output a predicted code ($abstract_p$). The IDs that the $abstract_p$ contains can be mapped back to real values only if they also appear in the input code. If the mutated code suggests to introduce a method call, \texttt{METHOD\_6}, which is not found in the input code, we cannot automatically map \texttt{METHOD\_6} to an actual method name. This inability to map back source code exists for any newly created ID generated for identifiers or literals, which are absent in the input code. Synthesizing new identifiers would involve extensive knowledge about the project, control and data flow information. For this reason, we discard the TPs that contain, in the buggy method $m_b$, new identifiers not seen in the fixed method $m_f$. The rationale is that we want to train our model from examples that rearrange existing identifiers, keywords and operators to mutate source code. Instead, this is not the case for literals. While we cannot perfectly map a new literal ID to a concrete value, we can still synthesize new literals by leveraging the type information embedded in the ID. For example, the (fixed) if condition in Fig. \ref{fig:tp} \texttt{if(VAR\_1.METHOD\_2( )  $<$ INT\_1)} should be mutated in its buggy version \texttt{if(VAR\_1.METHOD\_2( )  $<$ INT\_2)}. The value of \texttt{INT\_2} has never appeared in the input code (fixed), but we could still generate a compilable mutant by randomly generating a new integer value (different from any literal in the input code). While in these cases the literal value is randomly generated, the mutation model still provides the prediction about which literal to mutate.

For such reasons, we create two sets of TPs, hereby referred as $TP_{ident}$  and $TP_{ident-lit}$.
$TP_{ident}$ contains all TPs $tpa = \{abstract_b, abstract_f, A, M\}$ such that every identifier ID (\texttt{VAR\_}, \texttt{METHOD\_}, \texttt{TYPE\_}) in $abstract_b$ is found also in $abstract_f$. In this set we do allow new literal IDs (\texttt{STRING\_}, \texttt{INT\_}, \etc). $TP_{ident-lit}$ is a subset of $TP_{ident}$, which is more restrictive, and only contains the transformation pairs $tpa = \{abstract_b,$  $abstract_f, A, M\}$ such that every identifier and literal ID in $abstract_b$ is found also in $abstract_f$. Therefore, we do not allow new identifiers nor literals.

The rationale behind this choice is that we want to learn from examples (TPs) where the model is able to generate compilable mutants (\ie generate actual source code, with real values for IDs). In the case of the $TP_{ident-lit}$ set, the model will learn from examples that do not introduce any new identifier and literal. This means that the model will likely generate code for which every literal and identifier can be mapped to actual values. From the set $TP_{ident}$ the model will likely generate code for which we can map every identifier but we may need to generate new random literals.

In this context it is important to understand the role played by the idioms in our code representation. Idioms help  to retain transformation pairs that we would otherwise discard, and learn transformation of literal values that we would otherwise need to randomly generate. Consider again the previous example \texttt{if(VAR\_1 . METHOD\_2 ( )  < INT\_1)} and its mutated version \texttt{if(VAR\_1 . METHOD\_2 ( )  < INT\_2)}.  In this example, there are no idioms and, therefore, the model learns to mutate \texttt{INT\_1} to \texttt{INT\_2} within the \texttt{if} condition. However, when we want to map back the mutated (buggy) representation to actual source code, we will not have a value for  \texttt{INT\_2} (which does not appear in the input code) and, thus, we will be forced to generate a synthetic value for it. Instead, with the idiomized abstract representation the model would treat the idioms \texttt{0} and \texttt{1} as keywords of the language and learn the exact transformation of the \texttt{if} condition. The proposed mutant will therefore contain directly the idiom value (1) rather than \texttt{INT\_2}. Thus, the model will learn and propose such transformation without the need to randomly generate literal values. In summary, idioms increase the number of transformations incorporating real values rather than abstract representations. 
Without idioms, we would lose these transformations and our model could be less expressive. 

\subsubsection{Clustering}
\label{sec:clustering}
The goal of clustering is to create subsets of TPs such that TPs in each cluster share a similar list of AST actions. Each cluster represents a cohesive set of examples so that a trained a model can apply those actions to a new code. 

As previously explained, each transformation pair $tp = \{m_b, m_f, A\}$ includes a list of AST actions $A$. In our dataset, we found $\mathtt{\sim}$1,200 unique AST actions, and each TP can perform a different number and combination of these actions. Deciding whether the transformation pairs, $tp_1$ and $tp_2$, perform a similar sequence of actions and, thus, should be clustered together, is far from trivial. Possible similarity functions include the number of shared elements in the two sets of actions and the frequency of particular actions within the sets. Rather than defining such handcrafted rules, we choose to learn similarities directly from the data. We rely on an unsupervised learning algorithm that learns vector representations for the lists of actions $A$ of each TP. We treat each list of AST actions ($A$) as a document and rely on {\em doc2vec} \cite{Rong:2014} to learn a fixed-size vector representation of such variable-length documents embedded in a latent space where similarities can be computed as distances. The closer two vectors are, the more similar the content of the two corresponding documents. In other words, we mapped the problem of clustering TPs to the problem of clustering continuous valued vectors. To this goal, we use $k$-means clustering, requiring the number of clusters ($k$) into which to partition the data upfront. When choosing $k$, we need to balance two conflicting factors: (i) maximize the number of clusters so that we can train several different mutation models and, as a consequence, apply different mutations to a given piece of code; and (ii) have enough training examples (TPs) in each cluster to make the learning possible. Regarding the first point, we target at least three mutation models. Concerning the second point, with the available TPs dataset we could reasonably train no more than six clusters, so that each of those contain enough TPs. Thus, we experiment on the dataset $TP_{ident-lit}$ with values of $k$ going from 3 to 6 at steps of 1 and we evaluate each clustering solution in terms of its Silhouette statistic \cite{citeulike:1117716, Kogan:2007}, a metric used to judge the quality of clustering. We found that $k=5$ generates the clusters with the best overall Silhouette values. We cluster the dataset $TP_{ident-lit}$ into clusters: $C_1, C_2, C_3, C_4, C_5$.

\subsection{Learning Mutations}
\label{sec:learning}

\subsubsection{Dataset Preparation}
Given a set of TPs (\ie $TP_{ident}$,  $TP_{ident-lit}$, $C_1, \dots , C_5$) we use the instances to train our Encoder-Decoder model. Given a $tpa = \{abstract_b, abstract_f, A, M\}$ we use only the pair ($abstract_f, abstract_b$) of fixed and buggy abstracted code for learning. No additional information about the possible mutation actions ($A$) is provided during the learning process to the model. The given set of TPs is randomly partitioned into: training (80\%), evaluation (10\%), and test (10\%) sets. Before the partitioning, we make sure to remove any duplicated pairs ($abstract_f, abstract_b$) to not bias the results (\ie same pair both in training and test set).

\subsubsection{Encoder-Decoder Model}
Our models are based on an RNN Encoder-Decoder architecture, commonly adopted in Machine Translation \cite{kalchbrenner-blunsom:2013:EMNLP, DBLP:journals/corr/SutskeverVL14, DBLP:journals/corr/ChoMGBSB14}.  This model comprises two major components: an RNN Encoder, which \textit{encodes} a sequence of terms \boldmath{$x$} into a vector representation, and an RNN Decoder, which \textit{decodes} the representation into another sequence of terms $y$. The model learns a conditional distribution over a (output) sequence conditioned on another (input) sequence of terms: \unboldmath$P(y_1,.., y_m | x_1,.., x_n)$, where $n$ and $m$ may differ. In our case, given an input sequence $\mathbf{x} = abstract_f = (x_1,.., x_n)$ and a target sequence $\mathbf{y} = abstract_b = (y_1,.., y_m)$, the model is trained to learn the conditional distribution:  $P(abstract_b | abstract_f) = P(y_1,.., y_m | x_1,.., x_n)$, where $x_i$ and $y_j$ are abstracted source tokens: Java keywords, separators, IDs, and idioms. 
The Encoder takes as input a sequence $\mathbf{x} = (x_1,.., x_n)$ and produces a sequence of states $\mathbf{h} = (h_1,.., h_n)$. We rely on a bi-directional RNN Encoder \cite{DBLP:journals/corr/BahdanauCB14} which is formed by a backward and forward RNNs, which are able to create representations taking into account both past and future inputs \cite{DBLP:journals/corr/BritzGLL17}. That is, each state $h_i$ represents the concatenation of the states produced by the two RNNs reading the sequence in a forward and backward fashion: $h_i = [ \overrightarrow{h_i}; \overleftarrow{h_i}] $. The RNN Decoder predicts the probability of a target sequence $\mathbf{y} = (y_1,.., y_m)$ given $\mathbf{h}$. Specifically, the probability of each output term $y_i$ is computed based on: (i) the recurrent state $s_i$ in the Decoder; (ii) the previous $i-1$ terms $(y_1,.., y_{i-1})$; and (iii) a context vector $c_i$.  The latter practically constitutes the attention mechanism. The vector $c_i$ is computed as a weighted average of the states in $\mathbf{h}$, as follows: $c_i = \sum_{t=1}^{n}{a_{it}h_t}$ where the weights $a_{it}$ allow the model to pay more \textit{attention} to different parts of the input sequence. Specifically, the weight $a_{it}$ defines how much the term $x_i$ should be taken into account when predicting the target term $y_t$. The entire model is trained end-to-end (Encoder and Decoder jointly) by minimizing the negative log likelihood of the target terms, using stochastic gradient descent. 

\subsubsection{Configuration and Tuning}
For the RNN Cells we tested both LSTM \cite{Hochreiter:1997:LSM:1246443.1246450} and GRU \cite{DBLP:journals/corr/ChoMGBSB14}, founding the latter to be slightly more accurate and faster to train. Before settling on the bi-directional Encoder, we tested the unidirectional Encoder (with and without reversing the input sequence), but we consistently found the bi-directional one yielding more accurate results. Bucketing and padding was used to deal with the variable length of the sequences. We tested several combinations of the number of layers (1,2,3,4) and units (256, 512). The configuration that best balanced performance and training time was the one with 1 layer encoder, 2 layer decoder both with 256 units. We train our models for 40k epochs, which represented our empirically-based sweet spot between training time and loss function improvements.  The evaluation step was performed every 1k epochs.

\section{Experimental Design}
\label{sec:exp_design}
% !TeX root = ../ms.tex

The evaluation has been performed on the dataset of bug fixes described in Sec. \ref{sec:design} and  answers three RQs.

\textbf{RQ1: Can we learn how to generate mutants from bug-fixes?}
RQ1 investigates the extent to which bug fixes can be used to learn and generate mutants. We train  models based on the two datasets: $TP_{ident}$ and $TP_{ident-lit}$. We refer to such models with the name general models ($GM_{ident}$, $GM_{ident-lit}$), because they are trained using TPs of each dataset without clustering.
 Each dataset is partitioned into 80\% training, 10\% validation, 10\% testing.

\underline{\emph{BLEU Score.}} The first performance metric we use is the Bilingual Evaluation Understudy (BLEU) score, a metric used to assess the quality of a machine-translated sentence \cite{PapineniRWZ02}. BLEU scores require reference text to generate a score, which indicates how similar the candidate and reference texts are. 
The candidate and reference texts are broken into n-grams and the algorithm determines how many n-grams of the candidate text appear in the reference text. We report the global BLEU score, which is the geometric mean of all n-grams up to four. To assess our mutant generation approach, we first compute the BLEU score between the abstracted fixed code ($abstract_f$) and the corresponding target buggy code. This BLEU score serves as our baseline for comparison. We compute the BLEU score between the predicted mutant ($abstract_p$) and the target ($abstract_b$). The higher the BLEU score, the more similar $abstract_p$ is to $abstract_b$, \ie the actual buggy code.  To fully understand how similar our prediction is to the real buggy code, we need to compare the BLEU score with our baseline.  Indeed,  the input code (\ie the fixed code) provided to our model can be considered by itself as a ``close translation'' of the buggy code, therefore, helping in achieving a high BLEU score. To avoid this bias, we compare the BLEU score between the fixed code and the buggy code (baseline) with the BLUE score obtained when comparing the predicted buggy code ($abstract_p$) to the actual buggy code ($abstract_b$). If the BLEU score between $abstract_p$ and $abstract_b$ is higher than that one between $abstract_f$ and $abstract_b$, it means that the model transforms the input code ($abstract_f$) into a mutant ($abstract_p$) that is closer to the buggy code ($abstract_b$) than it was before the mutation, \ie the mutation goes in the right direction. In the opposite case, the predicted code represents a translation that is further from the buggy code when compared to the original input. To assess whether the differences in BLEU scores between the baseline and the models are statistically significant, we employ a technique devised by Zhang \etal \cite{Zhang04interpretingbleu/nist}. Given the test set, we generate $m=2,000$ test sets by sampling with replacement from the original test set. Then, we evaluate the BLEU score on the $m$ test sets both for our model and the baseline. Next, we compute the $m$ deltas of the scores: $\delta_i = model_i - baseline_i$. Given the distribution of the deltas, we select the 95\% confidence interval (CI) (\ie from the $2.5^{th}$ percentile to the $97.5^{th}$ percentile). If the CI is completely above or below the zero (\eg $2.5^{th}$ percentile $>$ 0) then the differences are statistically significant.

\underline{\emph{Prediction Classification.}}
Given $abstract_f$, $abstract_b$ and $abstract_p$, we classify each prediction into one of the following categories: (i) perfect prediction if $abstract_p = abstract_b$ (the model converts the fixed code back to its buggy version, thus reintroducing the original bug); (ii) bad prediction if $abstract_p = abstract_f$ (the model was not able to mutate the code and returned the same input code); and (iii) mutated prediction if $abstract_p \neq abstract_b$ AND $abstract_p \neq abstract_f$ (the model mutated the code, but differently than the target buggy code). We report raw numbers and percentages of the predictions falling in the described categories. 

\underline{\emph{Syntactic Correctness.}}
To be effective, mutants need to be syntactically correct, allowing the project to be compiled and tested against the test suite. We evaluate whether the models' predictions are lexically and syntactically correct by means of a Java lexer and parser. Perfect predictions and bad predictions are already known to be syntactically correct since we established the correctness of the buggy and fixed code when extracting the TPs. The correctness of the predictions within the mutated prediction category is instead unknown. For this reason, we report both the overall percentage of syntactically correct predictions as well as  the mutated predictions. We do not assess the compilability of the code.

\underline{\emph{Token-based Operations.}}
We analyzed and classified models' predictions also based on their tokens' operations, classifying  the predictions into one of four categories: (i) insertion if \#tokens predictions $>$ \#tokens input; (ii) changes if \#tokens prediction = \#tokens input AND prediction $\neq$ input; (iii) deletions if \#tokens prediction $<$ \#tokens input; (iv) none if prediction = input. This analysis aims to understand whether the models are able to insert, change or delete tokens.
 
\underline{\emph{AST-based Operations.}}
Next, we focus on the mutated predictions. These are not perfect predictions, but we are interested in understanding whether the transformations performed by the models are somewhat similar to the transformations between the fixed and buggy code. 
In other words, we investigate whether the model performs AST actions similar to the ones needed to transform the input (fixed) code into the corresponding buggy code.
Given the input fixed code $abstract_f$, the corresponding buggy code $abstract_b$, and the predicted mutant $abstract_p$, we extract with GumTreeDiff the following lists of AST actions: $A_{f-b} = actions(abstract_f \rightarrow abstract_b)$ and $A_{f-p} = actions(abstract_f \rightarrow abstract_p)$. We then compare the two lists of actions, $A_{f-b}$ and $A_{f-p}$, to assess their similarities. We report the percentage of mutated predictions whose list of actions $A_{f-p}$ contains the same elements and frequency of those found in $A_{f-b}$. We also report the percentage of mutated predictions when only comparing their unique actions and disregarding their frequency. In those cases, the model performed the same list of actions but possibly in a different order, location or frequency than those which led to the perfect prediction (buggy code).

\textbf{RQ2: Can we train different mutation models?}
RQ2 evaluates the five models trained using the five clusters of TPs. For each model, we evaluate its performance on the corresponding 10\% test set using the same analyses discussed for RQ1. In addition, we evaluate whether models belonging to different clusters generate different mutants. To this aim, we first concatenate the test set of each cluster into a single test set. Then, we feed each input instance in the test set (fixed code) to each and every mutation model $M_1,.., M_5$, obtaining five mutant outputs. After that, we compute the number of unique mutants generated by the models. For each input, the number of unique mutants ranges from one to five depending on how many models generate the same mutation. We report the distribution of unique mutants generated by the models. 

\textbf{RQ3: What are the characteristics of the mutants generated by the models?}
RQ3 qualitatively assesses the generated mutants through manual analysis. We first discuss some of the perfect predictions found by the models. Then, we focus our attention on the mutated predictions (neither perfect nor bad predictions).
We randomly selected a statistically significant sample from the mutated predictions of each cluster-model and manually analyzed them. The manual evaluation assesses (i) whether the functional behavior of the generated mutant differs from the original input code; (ii) the types of mutation operations performed by the model in generating the mutant.

Three judges, among the authors performed the analysis, and each instance was independently evaluated by two of them. The judges were presented with the original input code and the mutated code. The judges defined the mutation operations types in an open-coding fashion. Also, they were required to indicate whether the performed mutation changed the code behavior. After the initial evaluation, two of the three judges met to discuss and resolve the conflicts (\ie disagreement in the functional behavior change or in the set of mutation operators assigned)
 in the evaluation results. 
 We report the distribution of the mutation operators applied by the different cluster-models, and  highlight the differences.

\section{Results}
\label{sec:results}
% !TeX root = ../ms.tex

\textbf{RQ1: Can we learn how to generate mutants?}

\underline{\emph{BLEU Score.}} The top part of Table \ref{tab:bleu} shows the BLEU scores obtained by the two general models and compared with the baseline. The rightmost column represents the $2.5^{th}$ percentile of the distribution of the deltas. Compared to the baseline, the models achieve a better BLEU score when mutating the source code w.r.t. the target buggy code.  The differences are statistically significant, and the $2.5^{th}$ percentile of the distribution of the deltas (+5.63 and +7.97), shows that the models' BLEU scores are significantly higher than those obtained by the baseline. The observed increase in BLEU score indicates that the code mutated by our approach ($abstract_p$) is more similar to the buggy code ($abstract_b$) than the input code ($abstract_f$). Thus, the injected mutations push the fixed code towards a ``buggy'' state, exactly what we expect from mutation operators. While our baseline is relatively simple, improvements of few BLEU score points have been treated as ``considerable'' in neural machine translation tasks \cite{DBLP:journals/corr/WuSCLNMKCGMKSJL16}. 

\begin{table}[t]
\vspace{-0.1cm}
%\vspace{-\baselineskip}
\scriptsize
\caption{BLEU Score}
\vspace{-0.1cm}
\label{tab:bleu}
\centering
\resizebox{1\linewidth}{!}{
\begin{tabular}{l|c|c|c}
\toprule
\multirow{2}{*}{Model} & $abstract_f$ - $abstract_b$ &  $abstract_p$ - $abstract_b$ & $2.5^{\textbf{th}}$ percentile\\
& (baseline) & (mutation) & $\delta = $mutation - baseline\\
\midrule
$GM_{ident}$ & 71.85 & 76.68 & +5.63 \\
$GM_{ident-lit}$ & 70.07 & 76.92 & +7.97 \\
\midrule
$M_{1}$ & 67.18 & 82.16 & +17.01\\
$M_{2}$ & 51.58 & 50.96 & +1.01\\
$M_{3}$ & 81.89 & 83.15 & +0.94\\
$M_{4}$ & 67.04 & 78.87 & +12.45\\
$M_{5}$ & 65.68 & 77.73 & +13.51\\
\bottomrule
\end{tabular}
}
\vspace{-2\baselineskip}
\end{table}

\underline{\emph{Prediction Classification.}}
Table \ref{tab:classification} shows the raw numbers and percentages of predictions falling into the three categories previously described (\ie perfect, mutated, and bad predictions). The $GM_{ident}$ generated 1,991 (17\%) perfect predictions whereas  $GM_{ident-lit}$ 2,132  (21\%) perfect predictions.
We fed into the trained model a fixed piece of code, which the model has never seen before, and the model was able to perfectly predict the buggy version of that code, \ie to replicate the original bug. No information about the type of mutation operations to perform nor mutation locations are provided to the model. The fixed code is its only input. It is also important to note that, for the perfect predictions of the  $GM_{ident-lit}$  model, we can transform the entire abstracted code to the actual source code by mapping each and every ID to their corresponding value. The perfect predictions generated by $GM_{ident}$ can be mapped to actual source code but, in some cases, we might need to randomly generate new literal values.

$GM_{ident}$ and $GM_{ident-lit}$ generate 6,020 (52\%) and 5,240 (52\%) mutated predictions, respectively. While these predictions do not match the actual buggy code, they still represent meaningful mutants. We analyze these predictions in terms of syntactic correctness and types of operations they perform.

Finally, $GM_{ident}$ and $GM_{ident-lit}$ are not able to mutate the source code in 3,548 (30\%) and 2,644 (26\%) cases, respectively. While the percentages are non-negligible, it is still worth noting that overall, in 69\% and 73\% of cases, the models are able to mutate the code. These instances of bad predictions can be seen as cases where the model is unsure of about how to properly mutate the code. There are different strategies that could be adopted to force the model to mutate the code (\eg penalize during training predictions that are equal to the input, modify the inference step, or using beam search and select the prediction that is not equal to the input). 

\begin{table}[t]
\vspace{-0.1cm}
%\vspace{-\baselineskip}
\scriptsize
\caption{Prediction Classification}
\vspace{-0.1cm}
\label{tab:classification}
\centering
\begin{tabular}{l|c|c|c|r}
\toprule
Model &  Perfect pred. &  Mutated pred. & Bad pred. & Total\\
\midrule
$GM_{ident}$ & 1,991 (17\%) & 6,020 (52\%) & 3,548 (31\%) & 11,559\\
$GM_{ident-lit}$  & 2,132 (21\%) & 5,240 (52\%) & 2,644 (27\%) & 10,016\\
\midrule
$M_{1}$ & 1,348 (45\%) & 1,500 (49\%) & 190 (6\%) & 3,038\\
$M_{2}$  & 65 (9\%) & 635 (91\%) & 1 (0\%) & 701\\
$M_{3}$ & 392 (13\%) & 967 (33\%) & 1,603 (54\%) & 2,962\\
$M_{4}$  & 721 (29\%) & 1,453 (57\%) & 358 (14\%) & 2,532\\
$M_{5}$ & 366 (34\%) & 681 (63\%) & 33 (3\%) & 1,080\\
\bottomrule
\end{tabular}
\end{table}

%\begin{table}[t]
%\vspace{\baselineskip}
%\vspace{-\baselineskip}
%\scriptsize
%\caption{Prediction Classification}
%\label{tab:classification}
%\centering
%\begin{tabular}{l|c|c|c|r}
%\toprule
%Model &  Perfect pred. &  Mutated pred. & Bad pred. & Total\\
%\midrule
%$GM_{ident}$ & 1,991 (17.22\%) & 6,020 (52.08\%) & 3,548 (30.70\%) & 11,559\\
%$GM_{ident-lit}$  & 2,132 (21.29\%) & 5,240 (52.32\%) & 2,644 (26.39\%) & 10,016\\
%\midrule
%$M_{1}$ & 1,348 (44.37\%) & 1,500 (49.37\%) & 190 (6.26\%) & 3,038\\
%$M_{2}$  & 65 (9.27\%) & 635 (90.58\%) & 1 (0.15\%) & 701\\
%$M_{3}$ & 392 (13.23\%) & 967 (32.65\%) & 1,603 (54.12\%) & 2,962\\
%$M_{4}$  & 721 (28.48\%) & 1,453 (57.39\%) & 358 (14.13\%) & 2,532\\
%$M_{5}$ & 366 (33.89\%) & 681 (63.06\%) & 33 (3.05\%) & 1,080\\
%\bottomrule
%\end{tabular}
%\vspace{-1.5\baselineskip}
%\end{table}

\underline{\emph{Syntactic Correctness.}}
Table \ref{tab:syntax} reports the percentage of syntactically correct predictions performed by the model. More than 98\% of the model predictions are lexically and syntactically correct. When focusing on mutated predictions, the syntactic correctness is still very high ($>$96\%). This indicates that the model is able to learn the correct syntax rules from the abstracted representation we use as input/output of the model. While we do not report statistics on the compilability of the mutants, we can assume that the $\sim$20\% perfect predictions generated by the models are compilable, since they correspond to actual buggy code that was committed to software repositories. This means that the compilability rate of the mutants generated by our models is at least around 20\%. This is a very conservative estimation that does not consider the mutated predictions. Brown \etal \cite{Brown:2017:CFW:3106237.3106280} achieved a compilability rate of 14\%. Moreover, ``the majority of failed compilations (64\%) arise from simple parsing errors'' \cite{Brown:2017:CFW:3106237.3106280}, whereas we achieve a better-estimated compilability and a high percentage of syntactically correct predictions. 

\begin{table}[t]
\vspace{-0.2cm}
\scriptsize
\caption{Syntactic Correctness}
\vspace{-0.1cm}
\label{tab:syntax}
\centering
\begin{tabular}{l|c|c}
\toprule
Model &Mutated pred. & Overall\\
\midrule
$GM_{ident}$ & 96.96\% & 98.42\%\\
$GM_{ident-lit}$ & 96.56\% & 98.20\%\\
\midrule
$M_{1}$ & 96.07\% & 98.06\%\\
$M_{2}$  &  95.12\% & 95.58\%\\
$M_{3}$ &  94.42\% & 98.18\%\\
$M_{4}$  &  95.25\% & 97.27\%\\
$M_{5}$ &  91.48\% & 94.63\%\\
\bottomrule
\end{tabular}
\vspace{-1.5\baselineskip}
\end{table}

\underline{\emph{Token-based Operations.}}
Table \ref{tab:tokens} shows the classification of predictions based on the token-based operations performed by the models. 
$GM_{ident}$ and $GM_{ident-lit}$  generated predictions that resulted in the insertion of tokens in 1\% of the cases, changed nodes in 5\% and 3\% of the cases, and deletion of tokens in 63\% and 69\%, respectively. While most of the predictions resulted in token deletions, it is important to highlight that our models are able to generate predictions that insert and change tokens. We investigated whether these results were in-line with the actual data, or whether this was due to a drawback of our learning. We found that the operations performed in the bug-fixes we collected are: 72\% insertion, 8\% deletion, and 20\% changes. This means that bug-fixes mostly tend to perform insert operations (\eg adding an $\mathtt{if}$ statement to check for an exceptional condition), which means that when learning to inject bugs by mutating the code, it is expected to observe a vast majority of delete operations (see \tabref{tab:tokens}). 

\begin{table}[t]
\vspace{-0.1cm}
\scriptsize
\caption{Token-based Operations}
\vspace{-0.1cm}
\label{tab:tokens}
\centering
\begin{tabular}{l|c|c|c|r}
\toprule
Model &  Insertion & Changes & Deletion & None\\
\midrule
$GM_{ident}$ & 97 (1\%) & 624 (5\%) & 7,290 (63\%) & 3,548 (31\%)\\
$GM_{ident-lit}$  & 125 (1\%) & 264 (3\%) & 6,983 (70\%) & 2,644 (26\%)\\
\midrule
$M_{1}$ & 11 (0\%) & 30 (1\%) & 2,807 (93\%) & 190 (6\%)\\
$M_{2}$  &  27 (4\%) & 11 (2\%) & 662 (94\%) & 1 (0\%)\\
$M_{3}$ &  42 (2\%) & 217 (7\%) & 1,100 (37\%) & 1,603 (54\%)\\
$M_{4}$  &  87 (3\%) & 123 (5\%) & 1,964 (78\%) & 358 (14\%)\\
$M_{5}$ &  25 (2\%) & 20 (2\%) & 1,002 (93\%) & 33 (3\%)\\
\bottomrule
\end{tabular}
\end{table}

%\begin{table}[t]
%\vspace{\baselineskip}
%\vspace{-\baselineskip}
%\scriptsize
%\caption{Token-based Operations}
%\label{tab:tokens}
%\centering
%\begin{tabular}{l|c|c|c|r}
%\toprule
%Model &  Insertion & Changes & Deletion & None\\
%\midrule
%$GM_{ident}$ & 97 (0.83\%) & 624 (5.40\%) & 7,290 (63.07\%) & 3,548 (30.70\%)\\
%$GM_{ident-lit}$  & 125 (1.25\%) & 264 (2.64\%) & 6,983 (69.72\%) & 2,644 (26.39\%)\\
%\midrule
%$M_{1}$ & 11 (0.36\%) & 30 (0.98\%) & 2,807 (92.40\%) & 190 (6.26\%)\\
%$M_{2}$  &  27 (3.85\%) & 11 (1.57\%) & 662 (94.44\%) & 1 (0.15\%)\\
%$M_{3}$ &  42 (1.42\%) & 217 (7.33\%) & 1,100 (37.13\%) & 1,603 (54.12\%)\\
%$M_{4}$  &  87 (3.44\%) & 123 (4.86\%) & 1,964 (77.57\%) & 358 (14.13\%)\\
%$M_{5}$ &  25 (2.31\%) & 20 (1.86\%) & 1,002 (92.78\%) & 33 (3.05\%)\\
%\bottomrule
%\end{tabular}
%\vspace{-1.5\baselineskip}
%\end{table}

\underline{\emph{AST-based Operations.}}
Table \ref{tab:ast} reports the percentage of mutated predictions that share the same set or list of operations that would have led to the actual buggy code. $GM_{ident}$ and $GM_{ident-lit}$ generate a significant amount of mutated predictions which perform the same set (16\% and 24\% respectively) or the same type and frequency (14\% and 21\%) of operations w.r.t. the buggy code. This shows that our  models can still generate mutated code that is similar to the actual buggy code.
 
\begin{table}[t]
\vspace{-0.1cm}
\scriptsize
\caption{AST-based Operations}
\vspace{-0.1cm}
\label{tab:ast}
\centering
\begin{tabular}{l|c|c}
\toprule
Model &Same Operation Set  & Same Operation List \\
\midrule
$GM_{ident}$ & 16.02\% & 13.90\%\\
$GM_{ident-lit}$ & 24.44\% & 21.90\%\\
\midrule
$M_{1}$ & 54.46\% & 48.66\%\\
$M_{2}$  &  11.18\% & 10.23\%\\
$M_{3}$ &  15.20\% & 14.27\%\\
$M_{4}$  &  31.65\% & 29.24\%\\
$M_{5}$ &  41.55\% & 37.44\%\\
\bottomrule
\end{tabular}
\vspace{-1\baselineskip}
\vspace{-0.1cm}
\end{table}

{\em \underline{\textbf{Summary for RQ$_{1}$}.}} Our models are able to learn from bug-fixes how to mutate source code. The general models generate mutants that perfectly correspond to the original buggy code in $\mathtt{\sim}$20\% of the cases. The mutants generated are mostly syntactically correct ($>$98\%) and with an estimated compilability rate of at least 20\%. 

\textbf{RQ2: Can we train different mutation models? }
We present the performance of the cluster models $M_{1}$,..,$M_{5}$ based on the metrics introduced before. Each model has been trained and evaluated on the corresponding cluster of TPs, with respective sizes of $C_{1}$ = 30,385, $C_{2}$ = 7,016, $C_{3}$ = 29,625, $C_{4}$ = 25,320, and $C_{5}$ = 10,798. \smallskip

\underline{\emph{BLEU Score.}}
Table \ref{tab:bleu} shows the BLEU scores obtained by the five models. The BLEU scores for these models (mutation column) are relatively high, between 77.73 and 83.15 (with exception of model $M_2$), meaning that the mutated code generated by such models is a very close translation of the actual buggy code. Looking at the distribution of the deltas, we can notice that all the $2.5^{th}$ percentiles are greater than zero, meaning that the models achieve a BLEU score which is statistically better than the baselines. Even in the case of $M_2$, for which the global BLEU score is slightly lower than the baseline when the comparison is performed over 2,000 random samples, it outperforms the baseline.

\underline{\emph{Prediction Classification.}}
Table \ref{tab:classification} shows the raw numbers and percentages of predictions falling in the three categories we defined. Model $M_1$ achieves the highest percentage of Perfect predictions (44\%), followed by model $M_5$ (33\%) and model $M_4$ (28\%). This means that, given a fixed code, it is very likely that at least one of the models would predict the actual buggy code, as well as other interesting mutants. At the same time, the percentages of Bad predictions decreased significantly (except for $M_3$) w.r.t. the general models.

\begin{figure*}[t]
\vspace{-0.35cm}
	\centering
	\includegraphics[width=0.8\linewidth]{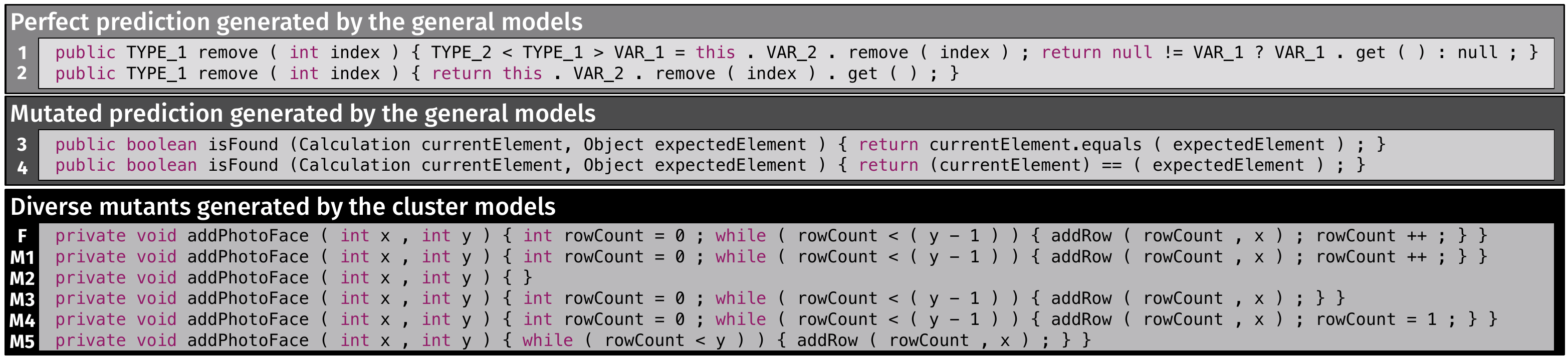}
	\vspace{-0.1cm}
	\caption{Qualitative Examples}
	\label{fig:qualitative}
	\vspace{-0.6cm}
\end{figure*}

The high percentage of bad predictions for $M_3$ can be partially explained by looking at the actual data in the cluster. The TPs in $C_3$ exhibits small transformations of the code. This is also noticeable from Table \ref{tab:bleu}, which shows a baseline BLEU score of 81.89 (the highest baseline value), which means that the input fixed code is already a close translation of the corresponding buggy code. This may have led the model to fall in a \textit{local minimum} where the mutation of the fixed code is the fixed code itself. Solutions for this problem may include: (i) further partitioning the cluster into more cohesive sub-clusters; (ii) allowing more training times/epochs for such models; (iii) implementing changes in the training/inference that we discussed previously.

\begin{figure}[t]
	\vspace{-0.1cm}
	\centering
	\includegraphics[width=0.7\linewidth]{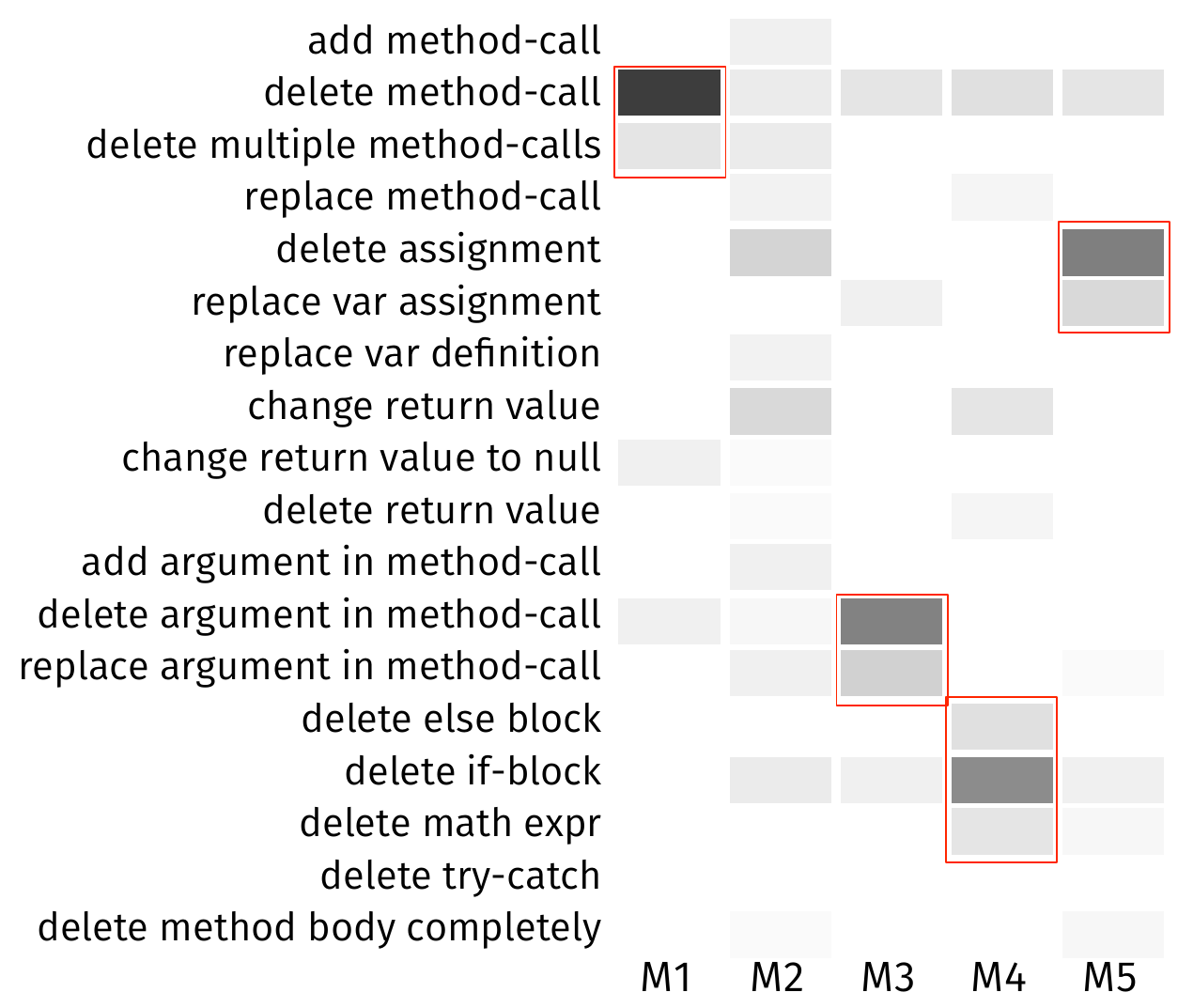}
	\vspace{-0.2cm}
	\caption{Cluster Models Operations}
	\label{fig:heatmap}
	\vspace{-0.7cm}
\end{figure}

\underline{\emph{Syntactic Correctness.}}
Table \ref{tab:syntax} reports the percentage of syntactically correct predictions performed by the models. Overall, the cluster model results are in-line with what was found for the general models, with an overall syntactic correctness between 94.63\% and 98.18\%. When focusing only on the mutated predictions, we still obtain very high syntactic correctness, between 91\% and 97\%. In terms of compilability, we could expect even better results for these models, given the higher rate of perfect predictions (which are likely to be compilable) generated by the cluster models. 

\underline{\emph{Token-based Operations.}}
Table \ref{tab:tokens} shows the classification of predictions based on the token-based operations performed by the models. The results for the cluster models are similar to what we found for general models, with higher percentages of deletions. In the next sections, we will look more into the differences among the operations performed by each model. 

\underline{\emph{AST-based Operations.}} 
Table \ref{tab:ast} reports the percentage of mutated predictions sharing the same set or list of operations w.r.t. the target buggy code. Cluster models, trained on cohesive sets of examples, generate a higher percentage of mutated predictions sharing the same set or list of operations as the target buggy code, as compared to the general models. The results for $M_1$, $M_4$, and $M_5$ are particularly good as they generate mutants with the same set of operations in 54\%, 31\%, and 41\% of the cases, respectively, and with the same list of operations in  48\%, 29\%, and 37\%, respectively.

\underline{\emph{Unique Mutants Generated.}}
The distribution of unique mutants generated by the five models has the 1st Qu. and Median values equal to 4, the mean equal to 4.2, and the 3rd Qu. equal to 5. Thus, the distribution appears to be skewed towards the maximum value (5). This demonstrates that we are able to train different mutation models that  generate diverse mutants given the same input code.

\underline{\emph{Generate Multiple Mutants.}}
We showed that clusters models are able to generate a diverse set of mutants, however it is also possible -- for each single model -- to generate $k$ different mutants for a given piece of code via beam search decoding. In a preliminary investigation we performed, we found that each model can generate more than 50 diverse mutants for a single method, with an impressive $\sim$80\% syntactic correctness.

{\em \underline{\textbf{Summary for RQ$_{2}$}.}}
The cluster models generate a high percentage of perfect predictions (between 9\% and 45\%) with  syntactic correctness between 94\% and 98\%. Even when the models generate mutants that are not perfect predictions, they usually apply a similar set of operations w.r.t. the buggy code. Furthermore, the trained models generate diverse mutants.

\textbf{RQ3: What are the characteristics of the mutants generated by the models? }
Fig. \ref{fig:qualitative} shows examples of perfect and mutated predictions generated by the general models, as well as diverse mutants generated by the cluster models for the same input code.  At the top, each example shows the input code (fixed) followed by the generated mutated code. 

The first example is a perfect prediction generated by the general model. The top line is the abstracted fixed code fed to the model, and the bottom line represents the output of the model, which perfectly corresponds to the target buggy code. The fixed code first removes the element at \texttt{index} from \texttt{VAR\_2}, assigning it to the \texttt{VAR\_1}, and then, if the newly defined variable is not null, it invokes the method \texttt{get} and returns its value, otherwise it returns null. The general model was able to apply different transformations of the code to generate the original buggy code, which invokes all the methods in sequence and returns the value.  If the removed element is null, the buggy code will throw an exception when invoking the method \texttt{get}. This transformation of the code does not fit in any existing mutation operator category.

Next, we report an interesting case of mutated prediction.  In this case, we used the mapping $M$ to automatically map back every identifier and literal, showing the ability to generate \textit{actual source code} from the output of the model. The model replaced the \texttt{equals()} method call with an equality expression (\texttt{==}). This shows how the model was able to learn common bugs introduced by developers when comparing objects. Note that the method name \texttt{equals} is an idiom, which allowed the model to learn this transformation.

Finally, the bottom part of Fig. \ref{fig:qualitative} shows the five mutants generated by the cluster models for the same fixed code (F) provided as input. In this case, we used the mapping $M$ to retrieve the source code from the output of the models. %We selected this example because it shows both interesting mutations and some limitations of our approach. 
$M_1$ was not able to generate a mutant and returned the same input code (bad prediction).  $M_2$ generated a mutant by removing the entire method body. While this appears like a trivial mutation, it is still meaningful as the method is not supposed to return a value, but only perform computations that will result in some side-effects in the class. This means that the test suite should carefully evaluate the state of the class after the invocation of the mutant. Mutants generated by $M_3$ and $M_4$ are the most interesting. They both introduce an infinite-loop, but in two different ways. $M_3$ deletes the increment of the \texttt{rowCount} variable, whereas $M_4$ resets its value to 1 at each iteration. Finally, $M_5$ changes the \texttt{if} condition and introduces an infinite loop similarly to the model $M_3$. However, it also deletes the variable definition statement for \texttt{rowCount}, making the mutant not compilable. 
All the predictions (including perfect, mutated, and diverse) are available in our appendix \cite{online-appendix}.

In the manual evaluation, three judges analyzed a total of 430 samples (90, 82, 86, 89, and 83 from $M_{1}$, $M_{2}$, $M_{3}$, $M_{4}$, $M_{5}$, respectively). 
In all cases except one, the judges agreed that the mutation code had a different behavior than the original code. The controversial case was related to a mutant which was created by deleting a print function call, which (only) indeed changed the method's output.

Fig. \ref{fig:heatmap} shows a heatmap of the frequency of mutation operations for each trained model. The intensity of the color represents the frequency with which a particular operation (specified by the row) was performed by the particular cluster model (columns). Due to space constraints, the rows of the heatmap contain only a subset of the 85 unique types of operations performed by the models, \ie only those performed in at least 5\% of the mutations by at least one model. 

We also highlighted in red boxes the peculiar, most frequent operations performed by each model. $M_1$ appears to focus on deletion of method calls; $M_3$ on deletion and replacement of an argument in a method call; $M_4$ mostly operates on \texttt{if-else} blocks and its logical conditions; $M_5$ focuses on deleting and replacing variable assignments. Finally, it is worth noting the large variety of operations performed by $M_2$, ranging from addition, deletion, and replacement of method calls, variable assignments, arguments, \etc. This might also explain the lower BLEU score achieved by the latter model, which performs large and more complex operations w.r.t. the other models which tend to focus on a smaller set of operations. Differences among the mutation models can also be appreciated by the number of different mutation operations performed for each mutant. The models $M_{1}$, $M_{2}$, $M_{3}$, $M_{4}$, $M_{5}$ performed  1.19, 3.48, 1.42, 1.93, 2.02 average number of operations for each mutant, respectively.

{\em \underline{\textbf{Summary for RQ$_{3}$}.}} The mutation models are capable of performing a diverse set of operations to mutate source code.

\section{Threats to Validity}
\label{sec:threats}
% !TeX root = ../ms.tex

\textbf{Construct validity.} To have enough training data, we mined bug-fixes in GitHub, rather than using curated datasets.
We disregarded large commits that might refer to tangled changes.

\textbf{Internal validity.} In assessing whether the generated mutants change the behavior of the code, we analyzed the mutated method in isolation (\ie not in the context of its system). This might have introduced imprecisions that were mitigated by assigning multiple evaluators to the analysis of each mutant. 

\textbf{External validity.} We only focused on Java code. However, the learning process is language-independent and can be instantiated for other languages by replacing the lexer, parser, and AST diff tools.
We only focused on methods having no more than 50 tokens. We also report experimental results on larger methods (50-100 tokens) using the same configuration of the network and training epochs \cite{online-appendix}. The model was still able to generate  $\mathtt{\sim}$6\% of perfect predictions. More training time and parameters' tuning can lead to better results.

\section{Related Work}
\label{sec:related}
% !TeX root = ../ms.tex

Brown \etal \cite{Brown:2017:CFW:3106237.3106280} leveraged bug-fixes to extract syntactic-mutation patterns from the diffs of patches. Our approach is novel and differs from Brown \etal in several aspects:

\begin{itemize}[noitemsep,wide=0pt, leftmargin=\dimexpr\labelwidth + 0\labelsep\relax]
	\item Rather than extracting all possible mutation operators from syntactic diffs, we automatically learn mutations from the data;
	\item Rather than focusing, in isolation, on contiguous lines of code changed in the diff, we are capable to learn which mutation operator is effective in a given  context;
	\item Our approach can automatically mutate identifiers and literal by inserting idioms in the new mutant. When the model suggests to mutate a literal with another unknown literal, it is generated randomly. Brown's \etal approach does not contemplate the synthesis of new identifiers (see Sec. 2.3 \cite{Brown:2017:CFW:3106237.3106280});
	\item Rather than extracting a single mutation pattern, we can learn co-occurrences and combinations of multiple mutations; 
	\item While the approach by Brown \etal randomly applies mutation operators to any code location unless the user specifies a rule for that, our approach automatically applies, for a given piece of code, the mutation(s) that according to the learning might reflect likely bugs occurring in such a location. While limiting mutants only to the most suitable ones for each location might not be always necessary, because one can apply as many mutants as possible to increase fault detection, this could lead to an overestimate of a test suite effectiveness or to more effort to unnecessarily augment a test suite. In a view of test suite optimization, an approach that learns where and how to mutate code is therefore desirable.
\end{itemize}

Different general-purpose mutation frameworks have been defined in the literature, including $\mu$Java~\cite{Ma:2005},  Jester~\cite{Jester},  Major~\cite{Just:ISSTA14}, Jumble~\cite{Jumble}, PIT~\cite{PIT}, and javaLanche~\cite{Schuler:FSE09}.
The main novelty of our work over those approaches is the automation of the learning and application of the mutation.

Relevant to our work are also studies investigating the relationship between mutants and real faults. Andrews \etal~\cite{AndrewsBL05,AndrewsBLN06} showed that carefully selected mutants can provide a good assessment of a test suite's ability to catch real faults and hand-seeded faults can underestimate the test suite's bug detection capability. Daran and Th\'evenod-Fosse \cite{DaranT96} found that the set of errors produced by carefully selected mutants and real faults with a given test suite are quite similar, while Just \etal \cite{Just:FSE14} reported that some types of real faults are not coupled to mutants and highlighted the need for new mutation operators. Chekham \etal \cite{Chekam:ICSE'17} showed that strong mutation testing yields high fault revelation, while this is not the case for weak mutation testing. Our work builds on these studies: we avoid the need for manually selecting the mutants, to increase the chances of generating mutants representative of real bugs.

Allamanis \etal~\cite{DBLP:journals/corr/AllamanisBJS16} generated tailored mutants, \eg exploiting API calls occurring elsewhere in the project and show that
 tailored mutants are well-coupled to real bugs. Differently from them, we automatically learn how to mutate code from an existing dataset of bugs rather than using heuristics.

Tufano \etal proposed the use of NMT for automated program repair \cite{TufanoWBPWP18,abs-1812-08693}. While mutation also requires learning from existing fixes, learning mutants is different because (i) it requires to abstract recurring mutants from multiple fixes (Sec. \ref{sec:clustering}) rather than learning and proposing specific fixes, and (ii) the learning happens in the opposite direction.

\section{Conclusion}
\label{sec:conclusion}
% !TeX root = ../ms.tex

We presented the first NMT-based approach to automatically learn mutants from existing bug fixes. The generated mutants are similar to real bugs, with 9\% to 45\% of them (depending on the model) reintroducing in the fixed code (provided as an input) the actual bug. Moreover, our models are able to learn the \textit{correct} code syntax, without the need for syntax rules as input. We release data and source code, so that researchers can use them for learning other transformations of code \cite{online-appendix}. Future work includes (i) additional fine-tuning of the RNN's parameters, and (ii) the creation of a mutation testing tool built on top of our approach. Indeed, while we achieved promising results in the challenging task of automatically learning mutants, it is worth noting that a whole infrastructure must be developed to automatically inject the learned mutants, making our approach usable by software developers.

\section{Acknowledgment}
Bavota acknowledges the support of the Swiss National Science Foundation for the CCQR project (No. 175513). This work is supported in part by the NSF CCF-1525902, CCF-1815186 and CCF-1927679 grants.

\bibliographystyle{IEEEtran}
\bibliography{IEEEabrv,ms}

\end{document}